\documentclass[pra,aps,reprint,superscriptaddress,longbibliography]{revtex4-1}
\usepackage[T1]{fontenc}
\usepackage{graphicx}
\usepackage{epstopdf}
\usepackage{amsmath,amssymb,amsfonts}
\usepackage{hyperref}
\usepackage[all]{hypcap}
\DeclareMathOperator{\e}{e}
\DeclareMathOperator{\ii}{i}
\makeatletter
\newcommand*\nobreakhyphen{\hbox{-}\nobreak\hskip\z@skip}
\makeatother
\begin{document}
\title{Robust stimulated Raman exact passage using shaped pulses}
\author{X.~Laforgue}\email{xavier-jacques_laforgue-marin@etu.u-bourgogne.fr}
\affiliation{\footnotesize Laboratoire Interdisciplinaire Carnot de Bourgogne, Universit\'e Bourgogne Franche-Comt\'e, CNRS UMR 6303, Alain Savary 9, 21078 Dijon, France}
\affiliation{\footnotesize Institut f\"ur Angewandte Physik, Technische Universit\"at Darmstadt, Hochschulstra\ss e 6, 64289 Darmstadt, Germany}
\author{Xi Chen}
\affiliation{\footnotesize Department of Physics, Shanghai University, 200444 Shanghai, People's Republic of China}
\author{S.~Gu\'erin}\email{sguerin@u-bourgogne.fr}
\affiliation{\footnotesize Laboratoire Interdisciplinaire Carnot de Bourgogne, Universit\'e Bourgogne Franche-Comt\'e, CNRS UMR 6303, Alain Savary 9, 21078 Dijon, France}
\date{\today}
\begin{abstract}
We developed single-shot shaped pulses for ultra high fidelity (UH-fidelity) population transfer on a 3-level quantum system in lambda configuration. To ensure high fidelity, we use the Lewis-Riesenfeld (L-R) method to derive a family of solutions leading to an exact transfer, where the solutions follow a single dynamical mode of the L-R invariant. Among this family, we identify a tracking solution with a single parameter to control simultaneously the fidelity of the transfer, the population of the excited state, and robustness. We define a measure of the robustness of an UH-fidelity transfer as the minimum percentile deviation on the pulse areas at which the infidelity rises above $10^{-4}$. The robustness of our shaped pulses is found superior to that of Gaussian and adiabatically-optimized pulses for moderate pulse areas.
\end{abstract}
\maketitle
\section{Introduction}
Three-level lambda ($\Lambda-$)systems are ubiquitous in quantum processes \cite{Gaubatz1990,Kuhn2002,Sorensen2006,Kral2007,Bergmann2015,Vitanov2017}. Many applications, particularly in quantum control, are based on our ability of controlling the population transfer in this system \cite{Gaubatz1990,Kuhn2002,Sorensen2006,Kral2007,Shore2011}. Fine control compatible with quantum information requirements imposes producing robust transfer of population with ultra high fidelity (UH-fidelity), i.e.~under the quantum computation infidelity benchmark of $\epsilon<10^{-4}$ \cite{Preskill1998}, between the two ground states of the system, while maintaining a low transient population on the intermediate (and often lossy) excited state during the dynamics. A standard method to perform this task is the stimulated Raman adiabatic passage, commonly known as STIRAP; widely used with applications in many physical and chemical problems \cite{Gaubatz1990,Kuhn2002,Sorensen2006,Kral2007,Shore2011,Bergmann2015,Vitanov2017}. 
STIRAP uses adiabaticity in order to avoid populating the intermediate state of a three-level system and to produce a robust transfer, at the expense of the process duration and pulse energy. It requires two fields: one, coupling the initial state of the system with the excited state, to which we refer as the pump $P$ and another, coupling the excited state with the target state, to which we refer as the Stokes field $S$; both names kept for historical reasons. In STIRAP, the fields coupling the ground states with the excited one must be counter-intuitively ordered (Stokes before pump) and exhibit high pulse areas and/or long time durations (technically any combination of factors fulfilling the adiabatic condition). Pulsed fields with increasingly higher areas and a counter-intuitive order, signatures of STIRAP, jointly with an optimized delay between the pulses, improve the adiabaticity and, in consequence, the robustness of the process, while minimizing the unwanted transient population of the excited state.

Even though STIRAP is the `go-to' standard protocol when to increase the process robustness becomes necessary, it is only at the adiabatic limit that it produces a complete transfer to the desired state and maintains the excited state depopulated.  That is to say that the target state $|\psi_T\rangle$ is approached asymptotically by the system state $|\psi(t)\rangle$ while the pulses areas $A_P=\int_{-\infty}^{\infty}P(t)dt$ and $A_S=\int_{-\infty}^{\infty}S(t)dt$ grow without limit. Concretely, the precision of the transfer can be measured with the fidelity $F=|\langle\psi_T|\psi(t_f)\rangle|^2=1-\epsilon$: a quantity equal to 1 when the transfer is perfect (target state achieved exactly by the system state at the process final time $t_f$) and to 0 when the final state is orthogonal to the target. Thus, in STIRAP, the fidelity tends to unity ($F\rightarrow1$) as the pulses areas tend to infinity ($\{A_P,A_S\}\rightarrow\infty$). In this manner, STIRAP provides a robust but inexact way of transferring population between the ground states of a three-level system. Additionally, the use of high area pulses hinders the application of such technique. Be it due to the destructive effects the usage of high intensity fields can produce, like ionization, or to the decoherence and experimental instabilities to which slow processes are susceptible, fields of moderate areas are most desirable for quantum state manipulation, especially for the UH-fidelity we aim at. Consequently, in this paper, we intend to propose a scheme for robust UH-fidelity transfers similar to STIRAP, but exact and with moderate areas, to which we refer as stimulated Raman exact passage (STIREP) \cite{Chen2012,Boscain2002,Dorier2017}. Exact, in this context, refers to schemes that provide the dynamics of the system ``exactly'', i.e.~approaches that prescribe a mathematical description of the complete dynamics of the system (the control fields and, consequently, the state are known for all instances of time).

Improvements of STIRAP have been proposed by optimizing single properties: nonresonant fast STIRAP \cite{Dridi2009} but with large transient population in the excited state and robust but slow STIRAP \cite{Vasilev2009}. However, there are exact methods available that take different approaches on their search to compete with STIRAP's well-stablished robustness, such as single-shot shaped pulses (SSSP) \cite{Daems2013,Van-Damme2017} and composite pulses \cite{Levitt1986,Torosov2011,Torosov2011a,Genov2011,Genov2014,Torosov2013,Bruns2018}, among others. Techniques as SSSP and composite pulses deal with error reduction directly, while methods as shortcuts to adiabaticity \cite{Demirplak2005,Demirplak2008,Dridi2009,Vasilev2009,Berry2009,Chen2010,Chen2011,Bason2012,Chen2012,Ruschhaupt2012,
Baksic2016,Li2016,Ban2018} rely on optimizing the adiabaticity of the process as their source of robustness. In a way, the first ones are bottom-up techniques, starting with energy economic strategies and remolding them to gain robustness; while the latter are top-down technologies, starting with the adiabatic and infinitely energetically costly paradigm and working their way down towards faster and cheaper processes. 

Physically speaking, exact methods are all those that offer detailed mathematical solutions for the desired task, i.e.~a description of the process with which to obtain the goal at a finite time. Meanwhile, adiabatic methods rely on the asymptotic behavior of the system under the adiabatic condition. To use an exact technique instead of an adiabatic one means to sacrifice the freedom that adiabaticity affords on field shapes for the rigidity of prescribed pulses and state dynamics. These prescriptions, provided by means of inverse engineering, are applied in order to gain the advantage of reaching the desired target state with finite pulse areas in a finite time.

SSSP is a technique that takes exact transfer reverse-engineering as a first step, and error resistance through the transfer perturbative expansion as a second step. Firstly, SSSP applies reverse-engineering from the desired process onto the control fields by means of the prescription of a tracking solution for a certain parameterization of the quantum state of the system. Then, it uses perturbation theory to gradually diminish the susceptibility of the transfer fidelity to deviations from the optimal experimental conditions. Perturbation theory is applied in terms of deviations from the ideal conditions, taking into consideration realistic experimental complications, and is analyzed through the Schr\"odinger equation. The minimization of the deviation terms, representing the result of non-optimal conditions, is expected to have the systematic decimation of the dynamics sensitivity to perturbations as a consequence, i.e.~improving the robustness. In order to manipulate the deviation terms, the tracking expression of the reverse-engineered dynamics must contain a suitable parameterization, meaning that the desired system evolution is prescribed with expressions containing free parameters to be chosen afterwards regarding they nullify or at least reduce the terms of the perturbative expansion.

In this paper, we introduce SSSP for the robust UH-fidelity transfer of population between the ground states of a 3-level $\Lambda$-system. We show a scheme similar to STIRAP but exact (thus not actually adiabatic) and highly robust using the Lewis-Riesenfeld (L-R) method driving a single dynamical mode \cite{Chen2011,Lewis1969}. The second section contains the parameterization of the propagator and Hamiltonian in terms of Euler angles. Section \ref{section2} shows the application of perturbation theory on the Hamiltonian, a working tracking solution (based on \cite{Boscain2002,Chen2012,Dorier2017}) and an analysis of the origin of robustness for this chosen tracking solution. We propose the direct study of the robustness of any given process for a range of pulse areas through the usage of a measurement of robustness based on the minimum UH-fidelity confidence range around the unperturbed ideal system. Additionally, definitions of STIRAP, considering Gaussian-shaped fields, and the adiabatically-optimized pulses with which we compare our SSSP are described. Section 4 presents the discussion and conclusions.
\section{\label{section1}The Hamiltonian and its state angular parameterization}
Let's consider a 3-level system driven by two resonant fields, $P(t)$ and $S(t)$, for which the Hamiltonian, on the bare states basis $\{|1\rangle,|2\rangle,|3\rangle\}$ and under the rotating wave approximation, is
\begin{equation}
\label{ResonantHamiltonian}
H(t) = \frac{\hbar}{2}\left[\begin{matrix}0&P&0\\P&0&S\\0&S&0\end{matrix}\right].
\end{equation}
In STIRAP, the state of the system is written in terms of the eigenstates of the Hamiltonian,
\begin{equation}
\label{darkstateSTIRAP}
\Phi_0=\left[\begin{matrix}\cos\vartheta\\0\\-\sin\vartheta\end{matrix}\right],\quad
\Phi_\pm=\frac1{\sqrt{2}}\left[\begin{matrix}\sin\vartheta\\\pm1\\\cos\vartheta\end{matrix}\right],
\end{equation}
where $\vartheta(t)$ is the so-called mixing angle, given by $\sin\vartheta=P/\sqrt{P^2+S^2}$ ($\cos\vartheta=S/\sqrt{P^2+S^2}$). The idea is to follow the dark state, the Hamiltonian eigenstate $|\Phi_0\rangle$, whose projection on the excited state is always null. This state allows for control of population transfer between the ground states without populating the intermediate state, the desired dynamics, which prescribes the signature counter-intuitive ordering of $P$ and $S$.
However, the derivatives of the mixing angle, the non-adiabatic coupling, couple the $|\Phi_n\rangle$'s, the adiabatic states, preventing their exact following (since population would be uncontrollably exchanged via it). Then, adiabaticity, the condition in which the non-adiabatic coupling is negligible (with $\dot{\vartheta}\rightarrow0$ being the adiabatic limit), is paramount to minimize the deviations of the dynamics from the dark state and produce the desired transfer.
Naturally, very slow-evolving pulses would minimize the non-adiabatic coupling and practically uncouple the adiabatic states in consequence. Nevertheless, the adiabatic states can never be followed exactly in real-world implementations.
\subsection{Lewis-Riesenfeld invariant}
A method that has taken notoriety in recent years is the use of dynamical invariants, also referred to as Lewis-Riesenfeld (L-R) invariants \cite{Lewis1969,Lai1996,Lai1996a,Chen2011,Chen2012}. 
The L-R invariant $I(t)$ is defined by having a time-invariant expectation value, i.e.~a constant $\langle\psi(t)|I|\psi(t)\rangle$, where $|\psi\rangle$ is the state of the system. This condition is equivalent to $\ii\hbar\dot{I}=[H,I]$ when considering the evolution of such system as described by the Schr\"odinger equation $\ii\hbar|\dot{\psi}(t)\rangle=H|\psi(t)\rangle$, where the dotted function denotes its partial derivative with respect to time.

We can use the eigenstates of this invariant, $|\varphi_n(t)\rangle$, to write the state of our system with the advantage that, unlike with the adiabatic states, the coupling between these is always null under any condition. 
This can be shown by applying the transformation operator $T_{\textrm{LR}}(t)=\sum_n|\varphi_n\rangle\langle n|$, that writes the system into the basis of the L-R eigenstates, onto the Sch\"odinger equation and demonstrating the effective Hamiltonian on the new basis, $H^{\textrm{LR}}(t)=T_{\textrm{LR}}HT_{\textrm{LR}}^\dagger-\ii\hbar T_{\textrm{LR}}\dot{T}_{\textrm{LR}}^\dagger$, to have only the diagonal elements $H^\textrm{LR}_n=\langle\varphi_n|H|\varphi_n\rangle$. Thus, we can describe the complete dynamics of our system by a fixed combination of the L-R eigenstates and, with a suitable parameterization and tracking solution, we can follow exactly the system evolution and, consequently, reach exactly the desired target state. 

A simple picture of the difference between the use of adiabatic states (key of STIRAP) and of the eigenvectors of the dynamical invariant (L-R method) is: while the adiabatic states represent the dynamics of the system under the adiabatic condition, the L-R eigenvectors contain the whole dynamics of the system; the firsts are a particular case of the seconds, as we will show at the end of this section.

In order to write the solution of the Schr\"odinger equation in terms of the eigenvectors of the L-R invariant we first need to write the latter explicitly in terms of practical parameters. For this purpose, we can exploit the property that establishes that, for an invariant that is member of the Lie algebra with (Hermitian) generators $Q_n$, i.e.~$I=\sum_n^N\alpha_n(t)Q_n$, these coefficients must obey the relation $\sum_n^N\alpha_n^2=\alpha_0^2$, where the $\alpha_n$'s are real quantities, $\alpha_0$ is a constant and $N$ is the number of generators of the algebra. 

Considering that the propagator of the Hamiltonian \eqref{ResonantHamiltonian} belongs to the $\mathrm{SU(3)}$ symmetry group, we can write said Hamiltonian as a linear combination of the well-known Gell-Mann matrices $\lambda_n$ of the group \cite{GellMann1962,Carroll1988,Chen2012} (generators of the Lie algebra of $\mathrm{SU(3)}$ as the Pauli matrices are the generators of the algebra of $\mathrm{SU(2)}$), i.e.~$H=\hbar/2(P\lambda_1+S\lambda_6)$, with
\begin{equation}
\lambda_1=\left[\begin{matrix}0&1&0\\1&0&0\\0&0&0\end{matrix}\right]\!,\ 
\lambda_5=\left[\begin{matrix}0&0&-\ii\\0&0&0\\\ii&0&0\end{matrix}\right]\!,\ 
\lambda_6=\left[\begin{matrix}0&0&0\\0&0&1\\0&1&0\end{matrix}\right].
\end{equation}
Moreover, given that the matrices $\lambda_1$, $\lambda_5$ and $\lambda_6$ form a closed algebra, fulfilling the Lie algebra of $\mathrm{SU(2)}$, i.e.~their commutation relations require no other generator ($[\lambda_i,\lambda_j]=C_{ij}^k\lambda_k$ for $i$, $j$, and $k$ taking any combination of values 1, 5 and 6 without repetitions, $C_{ij}^k=-C_{ji}^k=C_{jk}^i=C_{ki}^j$ and $C_{16}^5=\ii$), we can now write the L-R invariant in terms of only these three matrices and three $\alpha_n$'s:
\begin{equation}
I(t)=\alpha_1\lambda_1+\alpha_2\lambda_6+\alpha_3\lambda_5.
\end{equation}
This is a much simpler case than that of a general member of the $\mathrm{SU(3)}$ algebra that contains up to 8 $\alpha_n$'s (7 of which are independent). With this simple expression for our dynamical invariant we can solve the eigenvalue equation.

Using the eigenvectors $|\varphi_n(t)\rangle$ of this invariant to write the state of the system solution to the Schr\"odinger equation:
\begin{equation}
|\psi(t)\rangle=\sum_{n=1}^3C_n\e^{\ii\eta_n(t)}|\varphi_n(t)\rangle,
\end{equation}
with the Lewis-Riesenfeld phase 
\begin{equation}
\eta_n(t)=\frac1{\hbar}\int_{t_i}^t\left\langle\varphi_n(t')\left|\ii\hbar\frac{\partial}{\partial t'}-H(t')\right|\varphi_n(t')\right\rangle dt',
\end{equation}
we can also write the evolution operator $U(t,t_i)$, to which we refer as the propagator of the system, in terms of the $\alpha_n$'s through $U=\sum_{n=1}^3\exp[\ii\eta_n(t)]|\varphi_n(t)\rangle\langle\varphi_n(t_i)|$. The Lewis-Riesenfeld phase corresponding to the null eigenvalue, e.g., $\eta_1$, is a constant we set to 0. Considering we intend to prescribe the time evolution of the $\alpha_n$'s, we facilitate the search for the boundary conditions by imposing a single-mode driving, i.e.~a dynamics along a single eigenvector of the invariant, setting $C_1=1$ and $C_2=C_3=0$, which makes $|\psi\rangle=|\varphi_1\rangle$. This dynamics can be seen as a generalization of adiabatic passage, occurring along a single eigenstate, to an exact passage.

Given the relation between the $\alpha_n$'s, we can propose the following representation in terms of time-dependent Euler angles:
\begin{subequations}
\label{Euler-representation}
\begin{align}
\alpha_1&=\alpha_0\cos\phi\sin\theta,\\
\alpha_2&=-\alpha_0\cos\phi\cos\theta,\\
\alpha_3&=\alpha_0\sin\phi,
\end{align}
\end{subequations}
which consequently makes the other two L-R phases $\eta\equiv\eta_2=-\eta_3=-\int_{t_i}^t\dot{\theta}(t')/\sin[\phi(t')]dt'$. Defining the desired transfer to be $|\psi(t_i)\rangle=|1\rangle\rightarrow|\psi_T\rangle=|3\rangle$, we can now say that, for a Hamiltonian fulfilling the closed algebra of $\lambda_1$, $\lambda_5$ and $\lambda_6$, with no coupling $|1\rangle$--$|3\rangle$, the propagator of the system can be written as
\begin{equation}
U=\left[|\varphi_1\rangle\quad|\psi_+\rangle\quad|\psi_-\rangle\right],
\end{equation}
with the composing column vectors described by
\begingroup
\allowdisplaybreaks
\begin{subequations}
\label{propagatorvectors}
\begin{align}
\label{single-mode-eigenvector}
|\varphi_1\rangle&=\left[\begin{matrix}\cos\phi\cos\theta\\\ii\sin\phi\\\cos\phi\sin\theta\end{matrix}\right],\\
|\psi_+\rangle&=\left[\begin{matrix}\ii\cos\eta\sin\phi\cos\theta-\ii\sin\eta\sin\theta\\\cos\eta\cos\phi\\\ii\cos\eta\sin\phi\sin\theta+\ii\sin\eta\cos\theta\end{matrix}\right],\\
|\psi_-\rangle&=\left[\begin{matrix}-\sin\eta\sin\phi\cos\theta-\cos\eta\sin\theta\\\ii\sin\eta\cos\phi\\-\sin\eta\sin\phi\sin\theta+\cos\eta\cos\theta\end{matrix}\right],
\end{align}
\end{subequations}
\endgroup
where the first column of the propagator corresponds to a parameterization in Euler angles of the solution of the Schr\"odinger equation. With the representation in \eqref{Euler-representation}, the control fields can also be expressed in terms of these so-called Euler angles as
\begin{subequations}
\label{fields-Eulerangles}
\begin{align}
P/2&=-\dot{\theta}\cot\phi\sin\theta-\dot{\phi}\cos\theta,\\
S/2&=\dot{\theta}\cot\phi\cos\theta-\dot{\phi}\sin\theta,
\end{align}
\end{subequations}
which provide the remaining boundary conditions when demanding the pulses to have finite area, i.e.~$0\leftarrow P\rightarrow0$ and $0\leftarrow S\rightarrow0$, thus
\begin{equation}
\label{BoundaryConditons}
0\leftarrow\{\phi,\dot{\phi},\dot{\theta},\dot{\eta}\}\rightarrow0, \text{ and }0\leftarrow\theta\rightarrow\pi/2,
\end{equation}
where the arrows to the right and left represent the limits when $t\rightarrow t_f$ and $t\rightarrow t_i$, respectively.
It can be noted that the transient population of the excited state in this representation is given exactly by
\begin{equation}
\label{pop2-formula}
P_2(t)=|\langle2|\psi(t)\rangle|^2=\sin^2\phi(t).
\end{equation}

We can interpret the invariant's eigenstate $|\varphi_1\rangle$ as equivalent to the dark state of STIRAP, $|\Phi_0\rangle$, where the latter has been allowed to exhibit a non-zero transient excited state population in order to make the dynamics exact. In fact, the particular case of single-mode driving corresponding to adiabatic following is given by $|\Phi_0\rangle=|\varphi_1(\theta=-\vartheta,\phi=0)\rangle$; for which the excited state population \eqref{pop2-formula} remains exactly null, the fields \eqref{fields-Eulerangles} are infinite and, thus, the adiabatic condition is fulfilled.

Equations \eqref{fields-Eulerangles} and \eqref{BoundaryConditons} define a family of exact transfer solutions. Consequently, if such tracking solutions satisfying the previous conditions can be engineered, then we are able to control at will, in principle, the population on the middle state and we would be exposing an exact method for realizing stimulated Raman passage.
\section{\label{section2}Perturbed Hamiltonian, exact tracking and the measure of robustness}
Having set the requirements the angles must fulfill to describe the desired process, we proceed to deal with its robustness. Firstly, we add an unknown deviation $V(\rho)$ to the Hamiltonian \eqref{ResonantHamiltonian}, introducing the possibility of a non-optimal implementation of the control strategy that contains an error $\rho$ in the area of the pulses interacting with the system, i.e.~$H_\rho=H+V(\rho)$, where $V=\rho H$; thus,
\begin{equation}
\label{perturbedHamiltonian}
H_{\rho}=\frac{\hbar}{2}\left[\begin{matrix}0&(1+\rho)P&0\\(1+\rho)P&0&(1+\rho)S\\0&(1+\rho)S&0\end{matrix}\right].
\end{equation}
Secondly, we apply standard perturbation theory at the transfer profile regarding the perfect realization, or
\begin{subequations}
\label{transferprofile}
\begin{align}
\langle\psi_T|\psi_\rho(t_f)\rangle&=1+O_1+O_2+\cdots,\\
|\langle\psi_T|\psi_\rho(t_f)\rangle|^2&=1+\tilde{O}_1+\tilde{O}_2+\cdots=F.
\end{align}
\end{subequations}
The deviation terms $O_n\equiv O(\rho^n)$ are integral expressions whose level of complexity increases accordingly to the corresponding perturbation orders. Given that the evolution of the state of our system coincides with that of $|\varphi_1\rangle$, and that conjointly with $|\psi_+\rangle$ and $|\psi_-\rangle$ these form a complete basis, the deviation terms are, explicitly,
\begin{subequations}
\label{InfidelityIntegrals}
\begin{align}
O_1&=0,\\
O_2&=\int_{t_i}^{t_f}dt\int_{t_i}^tdt'\left[mm'-nn'\right]\in\Re,\\
O_3&=\int_{t_i}^{t_f}dt\int_{t_i}^tdt'\int_{t_i}^{t'}dt''[nr'm''-mr'n'']\in\Im,\\
O_4&=\int_{t_i}^{t_f}dt\int_{t_i}^tdt'\int_{t_i}^{t'}dt''\int_{t_i}^{t''}dt'''[mm'm''m'''\nonumber\\
&\quad-mm'n''n'''+mr'r''m'''-nn'm''m'''\nonumber\\
&\quad+nn'n''n'''-nr'r''n''']\in\Re,
\end{align}
\end{subequations}
and so on, where the non-null elements of the Hamiltonian deviation, for an unknown pulse area scaling error $\rho$, on the basis of the vectors in \eqref{propagatorvectors}, are identified as $m=\langle\varphi_1|V/\hbar|\psi_+\rangle$,  $n=\langle\varphi_1|V/\hbar|\psi_-\rangle$ and $r=\langle\psi_+|V/\hbar|\psi_-\rangle$, with the primed function representing the function with its argument primed, e.g., $m'=\langle\varphi_1(t')|V(t')/\hbar|\psi_+(t')\rangle$.

To consistently increase the robustness of the process via the nullification of the first orders of infidelity, $\tilde{O}_n\equiv\tilde{O}(\rho^n)$, is the goal of our strategy. These terms are, from \eqref{transferprofile}, given by $\tilde{O}_1=O_1+\bar{O}_1$, $\tilde{O}_2=O_2+O_1\bar{O}_1+\bar{O}_2$ and so on, where the odd orders are automatically null. However, the prescription of adequate tracking solutions with free parameters is the actual core of our recipe and also its sole non-systematic step.

Finally, we propose a tracking solution where the maximum transient population on the excited state, $P_2^{\mathrm{max}}=\mathrm{max}\left[|\langle2|\psi(t)\rangle|^2\right]$, is the control parameter. 
\subsection{\label{subsection1}Population cap parameterization}
The first found successful parameterization contains a unique free coefficient fixing a cap for the transient population on the excited state. The mixing angle of the levels $|1\rangle$ and $|3\rangle$ with $|2\rangle$, identified as $\phi(t)$, is written in terms of the other one, $\theta(t)$, which describes the state evolution from $|1\rangle$ to $|3\rangle$, and, in this manner, we propose the following suitable (fulfilling the requirements on \eqref{BoundaryConditons}) and convenient tracking solutions (based on \cite{Boscain2002,Chen2012,Dorier2017}):
\begin{subequations}
\begin{align}
\theta(t)&=(\pi/4)\left\{\tanh[(t-t_i-T/2)/v_0]+1\right\},\\
\tilde{\phi}(\theta)&=(4\phi_0/\pi)\sqrt{\theta(\pi/2-\theta)},
\end{align}
\end{subequations}
where the tilde signals functions of $\theta$. These give, with $\dot{\tilde{\phi}}\equiv\partial\tilde{\phi}/\partial\theta$,
\begin{subequations}
\begin{align}
\tilde{\dot{\theta}}(\theta)&=(4/\pi v_0)\theta(\pi/2-\theta),\\
\dot{\tilde{\phi}}(\theta)&=(4\phi_0/\pi)\frac{\pi/4-\theta}{\sqrt{\theta(\pi/2-\theta)}},
\end{align}
\end{subequations}
where $T=t_f-t_i$ is the total duration of the process and $v_0$ is a parameter setting the speed of the function change (chosen as $v_0=0.028T$ to provide a numerical error below $10^{-6}$ for the normalized field at the boundaries of the process).

\begin{figure}
\includegraphics[width=\columnwidth]{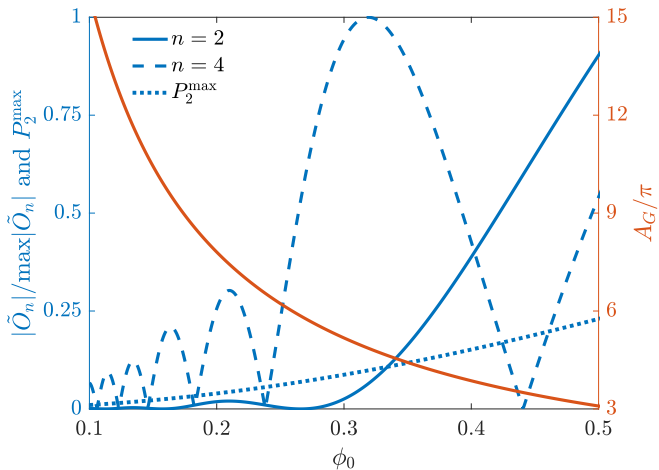}
\caption{\label{O2O4andAreaVSphi0}Second and fourth orders of infidelity, $\tilde{O}_2$ and $\tilde{O}_4$, maximum excited state population $P_2^{\mathrm{max}}$ and the corresponding generalized area $A_G$ vs the free parameter $\phi_0$.}
\end{figure}

The free parameter $\phi_0$ allows us to control simultaneously the maximum population on the excited state, parameterized as $P_2^{\mathrm{max}}=\sin^2\phi_0$, and the robustness of the transfer, by means of the nullification or minimization of the first orders of population infidelity $\tilde{O}_n$'s; the first two non-zero orders are shown in Fig.~\ref{O2O4andAreaVSphi0}.

The relationship between $\phi_0$ and the generalized area of the pulses, $A_G=\int_T\sqrt{P^2+S^2}dt$, corresponds to that which is well known from STIRAP: higher the area $A_G$ of the pulses, lower the maximum transient population on the excited state $P_2^{\mathrm{max}}$, which can be noted straightforwardly in Fig.~\ref{O2O4andAreaVSphi0}, to where we can also refer to extract the correspondence between $\phi_0$ and $A_G$. It can be highlighted that the additional amount of pulse area $\Delta A_G=A_G\left(P_2^{\mathrm{max}}(\phi_0)+\Delta P_2^{\mathrm{max}}\right)-A_G\left(P_2^{\mathrm{max}}(\phi_0)\right)$ that would be required to decrease the maximum intermediate state population by a certain amount $\Delta P_2^{\mathrm{max}}$ rises rapidly when considering ever lower values of $\phi_0$, i.e.~$\Delta A_G/\Delta P_2^{\mathrm{max}}\xrightarrow{\ \phi_0\rightarrow0\ }\infty$, thus exhibiting the asymptotic behavior of the adiabatic condition (the adiabatic limit).
\subsection{Measurement of robustness}
\subsubsection{Single-shot shaped pulses}
With the purpose of generating simple pulses, we choose to nullify the terms of the perturbative expansion of the infidelity maintaining a single control parameter. Since we only have one free variable, we can't, in general, use it to nullify more than one term; this is visible in Fig.~\ref{O2O4andAreaVSphi0}. However, given the particularity of our control, the absolute value of the perturbations, like the maximum population of the excited state, decreases in average as $\phi_0$ is decreased, contrary to the increase of the required pulse areas. We use this feature to restrict our focus to the range of $\phi_0$ corresponding to moderate pulse areas, e.g., $A_G\leq15\pi$, and examine the resultant robustness of the fidelity for the desired transfer.

\begin{figure}
\includegraphics[width=\columnwidth]{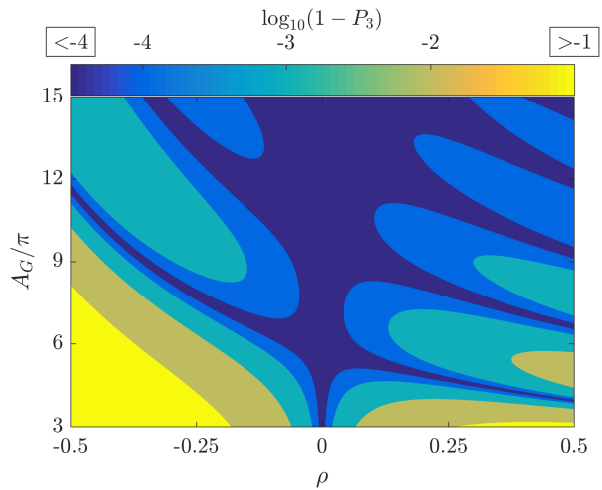}
\caption{\label{infidelityVSadiabAreaVSrho}Contour plot of infidelity $\epsilon$ (log base 10) vs generalized area $A_G$ and area perturbation $\rho$.}
\end{figure}

Considering the limited character of a single-parameter parameterization, we opt to not search to nullify individual terms of the perturbative expansion of the fidelity, but to search for particular values of $\phi_0$ for which the robustness of the transfer presents local maxima. Figure \ref{infidelityVSadiabAreaVSrho} permits to analyze the dependence of the infidelity $\epsilon$ on generalized pulse area and area scaling error $\rho$; this figure presents the contours of the regions with very high fidelities (over 99\%), showcasing them with the logarithm of the infidelity at the evaluated conditions, where we give special attention to the region of the so called ultra high fidelity (UH-fidelity) for which the infidelity $\epsilon\equiv1-P_3(t_f)\leq10^{-4}$.

The desired robustness can be understood as the non-susceptibility of the fidelity transfer (over a certain limit set to $10^{-4}$ for UH-fidelity) for different values of $\rho$, or how large does $\rho$ need to be (qualitatively around the unperturbed $\rho=0$ condition) to fall below the UH-fidelity definition. In Fig.~\ref{infidelityVSadiabAreaVSrho} we can observe how the robustness, in its qualitative sense from the broader UH-fidelity regions, tend to increase when more energy (or generalized pulse area) is invested.

The oscillatory behavior of the robustness is obtained from the oscillations of the infidelity orders $\tilde{O}_n$'s, shown in Fig.~\ref{O2O4andAreaVSphi0}, and the global increase of robustness with $A_G$ from the damping of such oscillations (the asymptotic decrease on the average of the absolute value of the infidelity orders). The asymmetry in Fig.~\ref{infidelityVSadiabAreaVSrho} arrives naturally from the fact that a positive $\rho$ increases the effective amplitude of the pulses, decreasing the generalized area required to achieve the UH-fidelity transfer, and vice versa.

\begin{figure}
\includegraphics[width=\columnwidth]{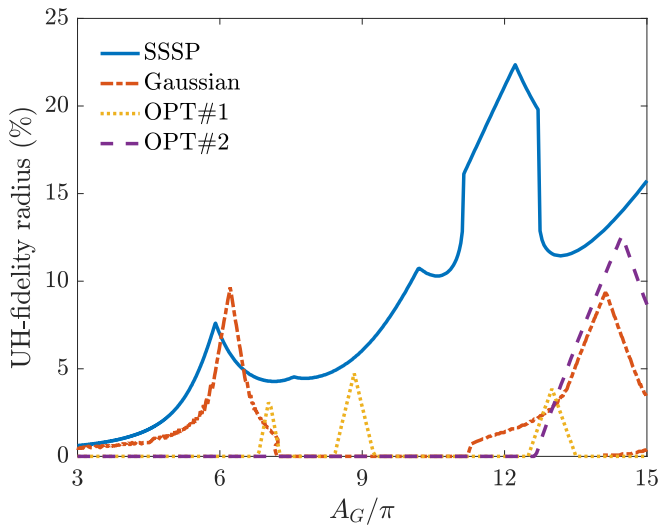}
\caption{\label{minDeltarhovsArea} UH-fidelity radius vs 
generalized area $A_G$. A comparison between selected techniques.}
\end{figure}

In order to have a quantitative measure of robustness, appropriate for its exhaustive analysis and for establishing grounds of comparison with other techniques, we extract the maximum absolute area deviation, $\max|\rho|$, at which transfers with ultra high fidelity are achieved for $\rho<0$ and $\rho>0$ separately. To the minimum of these two quantities we will refer as UH-fidelity radius and it is shown in Fig.~\ref{minDeltarhovsArea} in comparison with equal measures for Gaussian pulses and adiabatically-optimized pulses built from hypergaussians \citep{Vasilev2009}. We can remark that the discontinuous character of its definition, the operation of obtaining the minimum between the left and right values $|\rho|$ where the infidelity goes over $10^{-4}$, produces a UH-fidelity radius function with discontinuous derivatives.
\subsubsection{STIRAP with Gaussian pulses}
One of the most commonly used pulse shapes, especially for STIRAP, is Gaussian. Gaussian pulses have three free parameters: peak, waist and delay. The pulse areas $A_P$ and $A_S$ depend on the first two, and the generalized area $A_G$ depends on the three of them. Fixing the waist we can control the area by tuning the peak, but the efficiency of the process will also depend greatly on the delay. Thus, we optimize the delay and show the UH-fidelity radius in terms of $A_G$ to serve as a base reference for STIREP in Fig.~\ref{minDeltarhovsArea}.

For the Gaussian pulses, we use
\begin{subequations}
\begin{align}
P^{(G)}&=-\Upsilon\exp\left[-(\hat{t}-\tau/2)^2/\sigma^2\right],\\
S^{(G)}&=\Upsilon\exp\left[-(\hat{t}+\tau/2)^2/\sigma^2\right],
\end{align}
\end{subequations}
with $\hat{t}=t-t_i-T/2$. Where $\Upsilon$, $\tau$ and $\sigma$ are the peak, delay and waist of the gaussian pulses, respectively, which we restrict, while setting $\sigma=0.04T$, to a set of values that produce moderate area fields with smaller amplitudes (in their absolute values) than $10^{-6}\times\Upsilon$ at the boundaries of the process $[t_i,t_f]$, in order to have a proper numerical implementation with high precision.
\subsubsection{Adiabatically-optimized pulses}
The conditions for adiabatic optimization of pulse shapes, or designing adiabatically-optimal pulse shapes, are shown in \cite{Vasilev2009} while also proposing a combination of hypergaussian and trigonometric shapes as an example of pulses that fulfill these conditions for UH-fidelity STIRAP. The formulas for these pulses are:
\begin{subequations}
\begin{align}
P^{(O)}&=-\Upsilon\exp\left[-\left(\frac{\hat{t}}{m\sigma}\right)^{2n}\right]\sin\left(\frac{\pi/2}{f(\hat{t})}\right),\\
S^{(O)}&=\Upsilon\exp\left[-\left(\frac{\hat{t}}{m\sigma}\right)^{2n}\right]\cos\left(\frac{\pi/2}{f(\hat{t})}\right),
\end{align}
\end{subequations}
with $f=1+\exp(-\lambda\hat{t}/\sigma)$.
The dependence of the transfer robustness on area for a fixed waist $\sigma$, in order to be compared with Gaussian pulses of the same waist, has three remaining free parameters: $m$ (waist factor relative to the Gaussian pulses), $n$ (power of the hypergaussian) and $\lambda$ (speed of change of the trigonometric function).

These adiabatically-optimized pulses are shown \cite{Vasilev2009} to be superior to Gaussian pulses regarding the pulse area they require to achieve UH-fidelity standards when implemented for STIRAP. Moreover, these pulses are area-wise robuster than Gaussians when sufficiently (for UH-fidelity) high areas are used. The UH-fidelity radius of a pair of adiabatically-optimized pulses, labeled as OPT\#1 $(m=1,n=1,\lambda=4)$ and OPT\#2 $(m=1,n=2,\lambda=5)$ for two of the parameter sets (from sets with natural numbers as parameters) performing well at low to moderate pulse areas, is shown in Fig.~\ref{minDeltarhovsArea} for the purpose of comparison.
\section{Discussions and conclusion}
The UH-fidelity radius for the SSSP pulses developed in this paper, for Gaussian pulses and for adiabatically-optimized pulses is shown as a function of generalized area in Fig.~\ref{minDeltarhovsArea}.

SSSP is shown to be superior, for most areas under $A_G\leq15\pi$ at the very least, to the two other methods considered. The maximum of the UH-fidelity radius of SSSP is about 13\% over the Gaussian pulses with the highest performance and almost twice the maximum for the pulses OPT\#2, which is the second best performing technique, even though the latter requires over $2\pi$ higher pulse areas and is supposed to be, in that regard,  more adiabatic than the presented single-parameter SSSP.

Comparing Fig.~\ref{minDeltarhovsArea} with Fig.~\ref{O2O4andAreaVSphi0} we can discuss the locations of the maxima of the UH-fidelity radius for SSSP. From the low and insufficient pulse areas to the first maximum at about $6\pi$ we are observing the first minimum of the first non-null infidelity order $\tilde{O}_2$. The second most notable peak (neglecting the almost imperceptible one at $7.5\pi$) is located at about $10\pi$, an intermediate position between the second minimum of $\tilde{O}_2$ and the fourth of $\tilde{O}_4$. Finally, the largest, broadest and most relevant maxima to extract from this paper is located beyond the third minimum of $\tilde{O}_2$ and closer to, presumably, higher infidelity orders $\tilde{O}_n$'s. This UH-fidelity radius maxima at $\sim12\pi$ is the consequence of the simultaneous and local minimization of multiple infidelity orders and the best robustness obtained for $A_G\leq15\pi$ and among the comparable implementations of STIRAP shown on this study.

\begin{figure}
\includegraphics[width=\columnwidth]{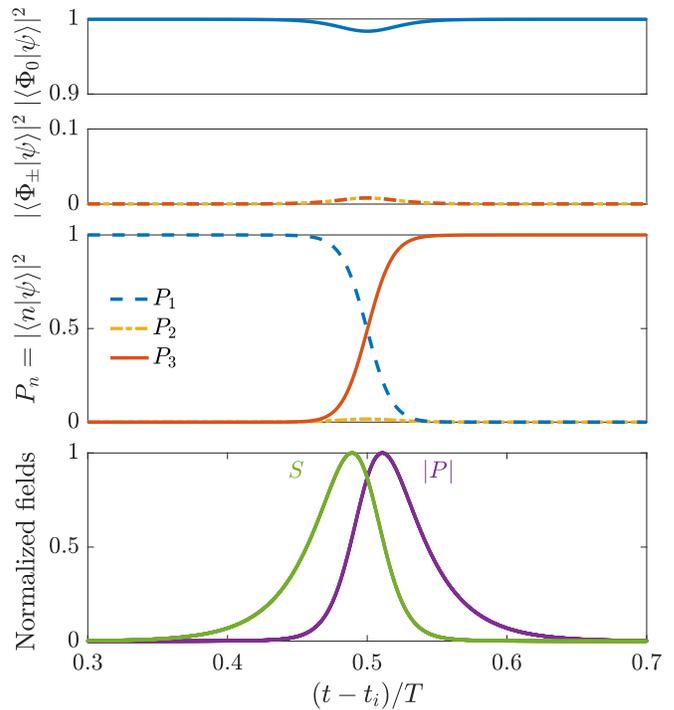}
\caption{\label{time-evolution} Time evolution of populations and the corresponding shaped fields, at best performing conditions, i.e.~$A_G\approx12\pi$ (regarding the UH-fidelity radius shown in Fig.~\ref{minDeltarhovsArea}).}
\end{figure}

The highest UH-fidelity radius reached by our SSSP, of 22.36\% for $A_G=12.23\pi$ or $\phi_0=0.12815$, generates the pulse shapes shown in Fig.~\ref{time-evolution} with its corresponding temporal population evolution and state's projection onto the adiabatic eigenvectors, time axis is limited to 40\% of the full time interval considered of duration $T$.

The projection of the state's dynamics onto the adiabatic states shows that the system doesn't follow the dark state along the evolution, it departs from it to populate a superposition of bright states, and, even though it comes back to it towards the end of the process, this differentiates it from the ideal STIRAP. In practice, this result would be similar for all counter-intuitively ordered control fields and differ only in the degree in which the excited state is populated during the dynamics. 

The pulses shapes are quite simple and similar to Gaussians but clearly asymmetric. The absolute value of the pump pulse, $|P|$, is shown instead of its direct value $P$, as it is shown for $S$, because observation is simplified this way, providing the figure with the only relevant information about the pulses: their shapes. 
For the same pulse shapes, pulses with equal or different relative signs will lead to identical results for the population fidelity; only the actual states involved would vary between $|1\rangle\rightarrow-|3\rangle$ (or $-|1\rangle\rightarrow|3\rangle$) for $P$ and $S$ of same sign, and $|1\rangle\rightarrow|3\rangle$ (or $-|1\rangle\rightarrow-|3\rangle$)
for $P$ and $S$ of different sign. The population of the excited state finds its maximum in the middle between the pulses, or $t-t_i=T/2$, and it has the reduced maximal value of $P_2=0.016$.

The UH-fidelity radius has been defined through the implementation of a Hamiltonian perturbation, shown in \eqref{perturbedHamiltonian}, that can be seen as considering a lack of perfect knowledge over the quantum system while having perfect control over the fields, some practical examples can be readily provided: 
\begin{itemize}
\item Pump and Stokes beams with equal intensity profiles (like Gaussian profiles with the same waist) interacting with atomic systems of no perfectly known location \cite{Bergmann1998}.
\item Certain variations on the dipole moment of the transitions, such as on their orientation, can affect both pump and Stokes fields on equal manner.
\item All those cases in which both controls are produced by the same source and thus any unexpected deviation affecting field amplitudes would be equal for the fields \cite{Vitanov2017}, such as when the considered transition frequencies are so close to each other that a single field can excite them. Another case would be that of when the addressed transitions involve Zeeman sublevels, where the coupling fields are only required to differ in polarization (right- and left-handed circular polarization for example). Having fields that originate from the same source impose them to have the same temporal shape, or to be mirror images of each other if we can use counter-propagating fields.
\end{itemize}
In conclusion, we have optimized robustness from an exact solution derived from the Lewis-Riesenfeld method with one mode, which allowed a full shaping of the fields. This strongly contrasts with respect to most of the previous attempts at optimizing STIRAP (fidelity and robustness) which were based on the optimization of a set of natural parameters, e.g., delay, waist, amplitude, among others. We have derived a parameterization achieving high robustness for moderate pulse areas. Additionally, this solution opens further prospects for designing various exact and robust solutions based on STIRAP, or its extensions, such as N-pod STIRAP \cite{Rousseaux2013} or other multilevel systems \cite{Hioe1985,Hioe1987,Hioe1988,Guengoerdue2012,Shore2013}.
\begin{acknowledgments}
This work was supported by the ``Investissements d'Avenir'' program, project ISITE-BFC / IQUINS (ANR-15-IDEX-03), and the EUR-EIPHI Graduate School (17-EURE-0002). X.L.~and S.G.~also acknowledge support from the European Union's Horizon 2020 research and innovation program under the Marie Sklodowska-Curie grant agreement No.~765075 (LIMQUET) and 641272 (HICONO). Also, X.C.~acknowledges the following funding:   NSFC (11474193),  the Shuguang Program (14SG35), the program of Shanghai Municipal Science and Technology Commission (18010500400 and 18ZR1415500). 
\end{acknowledgments}

\begin{thebibliography}{44}%
\makeatletter
\providecommand \@ifxundefined [1]{%
 \@ifx{#1\undefined}
}%
\providecommand \@ifnum [1]{%
 \ifnum #1\expandafter \@firstoftwo
 \else \expandafter \@secondoftwo
 \fi
}%
\providecommand \@ifx [1]{%
 \ifx #1\expandafter \@firstoftwo
 \else \expandafter \@secondoftwo
 \fi
}%
\providecommand \natexlab [1]{#1}%
\providecommand \enquote  [1]{``#1''}%
\providecommand \bibnamefont  [1]{#1}%
\providecommand \bibfnamefont [1]{#1}%
\providecommand \citenamefont [1]{#1}%
\providecommand \href@noop [0]{\@secondoftwo}%
\providecommand \href [0]{\begingroup \@sanitize@url \@href}%
\providecommand \@href[1]{\@@startlink{#1}\@@href}%
\providecommand \@@href[1]{\endgroup#1\@@endlink}%
\providecommand \@sanitize@url [0]{\catcode `\\12\catcode `\$12\catcode
  `\&12\catcode `\#12\catcode `\^12\catcode `\_12\catcode `\%12\relax}%
\providecommand \@@startlink[1]{}%
\providecommand \@@endlink[0]{}%
\providecommand \url  [0]{\begingroup\@sanitize@url \@url }%
\providecommand \@url [1]{\endgroup\@href {#1}{\urlprefix }}%
\providecommand \urlprefix  [0]{URL }%
\providecommand \Eprint [0]{\href }%
\providecommand \doibase [0]{https://doi.org/}%
\providecommand \selectlanguage [0]{\@gobble}%
\providecommand \bibinfo  [0]{\@secondoftwo}%
\providecommand \bibfield  [0]{\@secondoftwo}%
\providecommand \translation [1]{[#1]}%
\providecommand \BibitemOpen [0]{}%
\providecommand \bibitemStop [0]{}%
\providecommand \bibitemNoStop [0]{.\EOS\space}%
\providecommand \EOS [0]{\spacefactor3000\relax}%
\providecommand \BibitemShut  [1]{\csname bibitem#1\endcsname}%
\let\auto@bib@innerbib\@empty
\bibitem [{\citenamefont {Gaubatz}\ \emph {et~al.}(1990)\citenamefont
  {Gaubatz}, \citenamefont {Rudecki}, \citenamefont {Schiemann},\ and\
  \citenamefont {Bergmann}}]{Gaubatz1990}%
  \BibitemOpen
  \bibfield  {author} {\bibinfo {author} {\bibfnamefont {U.}~\bibnamefont
  {Gaubatz}}, \bibinfo {author} {\bibfnamefont {P.}~\bibnamefont {Rudecki}},
  \bibinfo {author} {\bibfnamefont {S.}~\bibnamefont {Schiemann}},\ and\
  \bibinfo {author} {\bibfnamefont {K.}~\bibnamefont {Bergmann}},\ }\bibfield
  {title} {\bibinfo {title} {{Population transfer between molecular vibrational
  levels by stimulated Raman scattering with partially overlapping laser
  fields. A new concept and experimental results}},\ }\href
  {https://doi.org/10.1063/1.458514} {\bibfield  {journal} {\bibinfo  {journal}
  {J.~Chem.~Phys.~}\ }\textbf {\bibinfo {volume} {92}},\ \bibinfo {pages}
  {5363} (\bibinfo {year} {1990})}\BibitemShut {NoStop}%
\bibitem [{\citenamefont {Kuhn}\ \emph {et~al.}(2002)\citenamefont {Kuhn},
  \citenamefont {Hennrich},\ and\ \citenamefont {Rempe}}]{Kuhn2002}%
  \BibitemOpen
  \bibfield  {author} {\bibinfo {author} {\bibfnamefont {A.}~\bibnamefont
  {Kuhn}}, \bibinfo {author} {\bibfnamefont {M.}~\bibnamefont {Hennrich}},\
  and\ \bibinfo {author} {\bibfnamefont {G.}~\bibnamefont {Rempe}},\ }\bibfield
   {title} {\bibinfo {title} {{Deterministic Single-Photon Source for
  Distributed Quantum Networking}},\ }\href
  {https://doi.org/10.1103/PhysRevLett.89.067901} {\bibfield  {journal}
  {\bibinfo  {journal} {Phys. Rev. Lett.}\ }\textbf {\bibinfo {volume} {89}},\
  \bibinfo {pages} {067901} (\bibinfo {year} {2002})}\BibitemShut {NoStop}%
\bibitem [{\citenamefont {S{\o}rensen}\ \emph {et~al.}(2006)\citenamefont
  {S{\o}rensen}, \citenamefont {M{\o}ller}, \citenamefont {Iversen},
  \citenamefont {Thomsen}, \citenamefont {Jensen}, \citenamefont {Staanum},
  \citenamefont {Voigt},\ and\ \citenamefont {Drewsen}}]{Sorensen2006}%
  \BibitemOpen
  \bibfield  {author} {\bibinfo {author} {\bibfnamefont {J.~L.}\ \bibnamefont
  {S{\o}rensen}}, \bibinfo {author} {\bibfnamefont {D.}~\bibnamefont
  {M{\o}ller}}, \bibinfo {author} {\bibfnamefont {T.}~\bibnamefont {Iversen}},
  \bibinfo {author} {\bibfnamefont {J.~B.}\ \bibnamefont {Thomsen}}, \bibinfo
  {author} {\bibfnamefont {F.}~\bibnamefont {Jensen}}, \bibinfo {author}
  {\bibfnamefont {P.}~\bibnamefont {Staanum}}, \bibinfo {author} {\bibfnamefont
  {D.}~\bibnamefont {Voigt}},\ and\ \bibinfo {author} {\bibfnamefont
  {M.}~\bibnamefont {Drewsen}},\ }\bibfield  {title} {\bibinfo {title}
  {{Efficient coherent internal state transfer in trapped ions using stimulated
  Raman adiabatic passage}},\ }\href
  {https://doi.org/10.1088/1367-2630/8/11/261} {\bibfield  {journal} {\bibinfo
  {journal} {New J. Phys.}\ }\textbf {\bibinfo {volume} {8}},\ \bibinfo {pages}
  {261} (\bibinfo {year} {2006})}\BibitemShut {NoStop}%
\bibitem [{\citenamefont {Kr{\'{a}}l}\ \emph {et~al.}(2007)\citenamefont
  {Kr{\'{a}}l}, \citenamefont {Thanopulos},\ and\ \citenamefont
  {Shapiro}}]{Kral2007}%
  \BibitemOpen
  \bibfield  {author} {\bibinfo {author} {\bibfnamefont {P.}~\bibnamefont
  {Kr{\'{a}}l}}, \bibinfo {author} {\bibfnamefont {I.}~\bibnamefont
  {Thanopulos}},\ and\ \bibinfo {author} {\bibfnamefont {M.}~\bibnamefont
  {Shapiro}},\ }\bibfield  {title} {\bibinfo {title} {{Colloquium: Coherently
  controlled adiabatic passage}},\ }\href
  {https://doi.org/10.1103/RevModPhys.79.53} {\bibfield  {journal} {\bibinfo
  {journal} {Rev. Mod. Phys.}\ }\textbf {\bibinfo {volume} {79}},\ \bibinfo
  {pages} {53} (\bibinfo {year} {2007})}\BibitemShut {NoStop}%
\bibitem [{\citenamefont {Bergmann}\ \emph {et~al.}(2015)\citenamefont
  {Bergmann}, \citenamefont {Vitanov},\ and\ \citenamefont
  {Shore}}]{Bergmann2015}%
  \BibitemOpen
  \bibfield  {author} {\bibinfo {author} {\bibfnamefont {K.}~\bibnamefont
  {Bergmann}}, \bibinfo {author} {\bibfnamefont {N.~V.}\ \bibnamefont
  {Vitanov}},\ and\ \bibinfo {author} {\bibfnamefont {B.~W.}\ \bibnamefont
  {Shore}},\ }\bibfield  {title} {\bibinfo {title} {{Perspective: Stimulated
  Raman adiabatic passage: The status after 25 years}},\ }\href
  {https://doi.org/10.1063/1.4916903} {\bibfield  {journal} {\bibinfo
  {journal} {J. Chem. Phys.}\ }\textbf {\bibinfo {volume} {142}},\ \bibinfo
  {pages} {170901} (\bibinfo {year} {2015})}\BibitemShut {NoStop}%
\bibitem [{\citenamefont {Vitanov}\ \emph {et~al.}(2017)\citenamefont
  {Vitanov}, \citenamefont {Rangelov}, \citenamefont {Shore},\ and\
  \citenamefont {Bergmann}}]{Vitanov2017}%
  \BibitemOpen
  \bibfield  {author} {\bibinfo {author} {\bibfnamefont {N.~V.}\ \bibnamefont
  {Vitanov}}, \bibinfo {author} {\bibfnamefont {A.~A.}\ \bibnamefont
  {Rangelov}}, \bibinfo {author} {\bibfnamefont {B.~W.}\ \bibnamefont
  {Shore}},\ and\ \bibinfo {author} {\bibfnamefont {K.}~\bibnamefont
  {Bergmann}},\ }\bibfield  {title} {\bibinfo {title} {Stimulated {R}aman
  adiabatic passage in physics, chemistry, and beyond},\ }\href
  {https://doi.org/10.1103/RevModPhys.89.015006} {\bibfield  {journal}
  {\bibinfo  {journal} {Rev. Mod. Phys.}\ }\textbf {\bibinfo {volume} {89}},\
  \bibinfo {pages} {015006} (\bibinfo {year} {2017})}\BibitemShut {NoStop}%
\bibitem [{\citenamefont {Shore}(2011)}]{Shore2011}%
  \BibitemOpen
  \bibfield  {author} {\bibinfo {author} {\bibfnamefont {B.~W.}\ \bibnamefont
  {Shore}},\ }\href@noop {} {\emph {\bibinfo {title} {Manipulating Quantum
  Structures Using Laser Pulses}}}\ (\bibinfo  {publisher} {Cambridge
  University Press},\ \bibinfo {year} {2011})\BibitemShut {NoStop}%
\bibitem [{\citenamefont {Preskill}(1998)}]{Preskill1998}%
  \BibitemOpen
  \bibfield  {author} {\bibinfo {author} {\bibfnamefont {J.}~\bibnamefont
  {Preskill}},\ }\bibfield  {title} {\bibinfo {title} {{Reliable quantum
  computers}},\ }\href {https://doi.org/10.1098/rspa.1998.0167} {\bibfield
  {journal} {\bibinfo  {journal} {Proc. R. Soc. Lond. A}\ }\textbf {\bibinfo
  {volume} {454}},\ \bibinfo {pages} {385} (\bibinfo {year}
  {1998})}\BibitemShut {NoStop}%
\bibitem [{\citenamefont {Chen}\ and\ \citenamefont {Muga}(2012)}]{Chen2012}%
  \BibitemOpen
  \bibfield  {author} {\bibinfo {author} {\bibfnamefont {X.}~\bibnamefont
  {Chen}}\ and\ \bibinfo {author} {\bibfnamefont {J.~G.}\ \bibnamefont
  {Muga}},\ }\bibfield  {title} {\bibinfo {title} {Engineering of fast
  population transfer in three-level systems},\ }\href
  {https://doi.org/10.1103/PhysRevA.86.033405} {\bibfield  {journal} {\bibinfo
  {journal} {Phys. Rev. A}\ }\textbf {\bibinfo {volume} {86}},\ \bibinfo
  {pages} {033405} (\bibinfo {year} {2012})}\BibitemShut {NoStop}%
\bibitem [{\citenamefont {Boscain}\ \emph {et~al.}(2002)\citenamefont
  {Boscain}, \citenamefont {Charlot}, \citenamefont {Gauthier}, \citenamefont
  {Gu\'erin},\ and\ \citenamefont {Jauslin}}]{Boscain2002}%
  \BibitemOpen
  \bibfield  {author} {\bibinfo {author} {\bibfnamefont {U.}~\bibnamefont
  {Boscain}}, \bibinfo {author} {\bibfnamefont {G.}~\bibnamefont {Charlot}},
  \bibinfo {author} {\bibfnamefont {J.~P.}\ \bibnamefont {Gauthier}}, \bibinfo
  {author} {\bibfnamefont {S.}~\bibnamefont {Gu\'erin}},\ and\ \bibinfo
  {author} {\bibfnamefont {H.~R.}\ \bibnamefont {Jauslin}},\ }\bibfield
  {title} {\bibinfo {title} {Optimal control in laser-induced population
  transfer for two- and three-level quantum systems},\ }\href
  {https://doi.org/10.1063/1.1465516} {\bibfield  {journal} {\bibinfo
  {journal} {J. Math. Phys.}\ }\textbf {\bibinfo {volume} {43}},\ \bibinfo
  {pages} {2107} (\bibinfo {year} {2002})}\BibitemShut {NoStop}%
\bibitem [{\citenamefont {Dorier}\ \emph {et~al.}(2017)\citenamefont {Dorier},
  \citenamefont {Gevorgyan}, \citenamefont {Ishkhanyan}, \citenamefont {Leroy},
  \citenamefont {Jauslin},\ and\ \citenamefont {Gu\'erin}}]{Dorier2017}%
  \BibitemOpen
  \bibfield  {author} {\bibinfo {author} {\bibfnamefont {V.}~\bibnamefont
  {Dorier}}, \bibinfo {author} {\bibfnamefont {M.}~\bibnamefont {Gevorgyan}},
  \bibinfo {author} {\bibfnamefont {A.}~\bibnamefont {Ishkhanyan}}, \bibinfo
  {author} {\bibfnamefont {C.}~\bibnamefont {Leroy}}, \bibinfo {author}
  {\bibfnamefont {H.~R.}\ \bibnamefont {Jauslin}},\ and\ \bibinfo {author}
  {\bibfnamefont {S.}~\bibnamefont {Gu\'erin}},\ }\bibfield  {title} {\bibinfo
  {title} {Nonlinear stimulated {R}aman exact passage by resonance-locked
  inverse engineering},\ }\href
  {https://doi.org/10.1103/PhysRevLett.119.243902} {\bibfield  {journal}
  {\bibinfo  {journal} {Phys. Rev. Lett.}\ }\textbf {\bibinfo {volume} {119}},\
  \bibinfo {pages} {243902} (\bibinfo {year} {2017})}\BibitemShut {NoStop}%
\bibitem [{\citenamefont {Dridi}\ \emph {et~al.}(2009)\citenamefont {Dridi},
  \citenamefont {Gu\'erin}, \citenamefont {Hakobyan}, \citenamefont {Jauslin},\
  and\ \citenamefont {Eleuch}}]{Dridi2009}%
  \BibitemOpen
  \bibfield  {author} {\bibinfo {author} {\bibfnamefont {G.}~\bibnamefont
  {Dridi}}, \bibinfo {author} {\bibfnamefont {S.}~\bibnamefont {Gu\'erin}},
  \bibinfo {author} {\bibfnamefont {V.}~\bibnamefont {Hakobyan}}, \bibinfo
  {author} {\bibfnamefont {H.~R.}\ \bibnamefont {Jauslin}},\ and\ \bibinfo
  {author} {\bibfnamefont {H.}~\bibnamefont {Eleuch}},\ }\bibfield  {title}
  {\bibinfo {title} {Ultrafast stimulated {R}aman parallel adiabatic passage by
  shaped pulses},\ }\href {https://doi.org/10.1103/PhysRevA.80.043408}
  {\bibfield  {journal} {\bibinfo  {journal} {Phys. Rev. A}\ }\textbf {\bibinfo
  {volume} {80}},\ \bibinfo {pages} {043408} (\bibinfo {year}
  {2009})}\BibitemShut {NoStop}%
\bibitem [{\citenamefont {Vasilev}\ \emph {et~al.}(2009)\citenamefont
  {Vasilev}, \citenamefont {Kuhn},\ and\ \citenamefont
  {Vitanov}}]{Vasilev2009}%
  \BibitemOpen
  \bibfield  {author} {\bibinfo {author} {\bibfnamefont {G.~S.}\ \bibnamefont
  {Vasilev}}, \bibinfo {author} {\bibfnamefont {A.}~\bibnamefont {Kuhn}},\ and\
  \bibinfo {author} {\bibfnamefont {N.~V.}\ \bibnamefont {Vitanov}},\
  }\bibfield  {title} {\bibinfo {title} {Optimum pulse shapes for stimulated
  {R}aman adiabatic passage},\ }\href
  {https://doi.org/10.1103/PhysRevA.80.013417} {\bibfield  {journal} {\bibinfo
  {journal} {Phys. Rev. A}\ }\textbf {\bibinfo {volume} {80}},\ \bibinfo
  {pages} {013417} (\bibinfo {year} {2009})}\BibitemShut {NoStop}%
\bibitem [{\citenamefont {Daems}\ \emph {et~al.}(2013)\citenamefont {Daems},
  \citenamefont {Ruschhaupt}, \citenamefont {Sugny},\ and\ \citenamefont
  {Gu\'erin}}]{Daems2013}%
  \BibitemOpen
  \bibfield  {author} {\bibinfo {author} {\bibfnamefont {D.}~\bibnamefont
  {Daems}}, \bibinfo {author} {\bibfnamefont {A.}~\bibnamefont {Ruschhaupt}},
  \bibinfo {author} {\bibfnamefont {D.}~\bibnamefont {Sugny}},\ and\ \bibinfo
  {author} {\bibfnamefont {S.}~\bibnamefont {Gu\'erin}},\ }\bibfield  {title}
  {\bibinfo {title} {Robust quantum control by a single-shot shaped pulse},\
  }\href {https://doi.org/10.1103/PhysRevLett.111.050404} {\bibfield  {journal}
  {\bibinfo  {journal} {Phys. Rev. Lett.}\ }\textbf {\bibinfo {volume} {111}},\
  \bibinfo {pages} {050404} (\bibinfo {year} {2013})}\BibitemShut {NoStop}%
\bibitem [{\citenamefont {Van-Damme}\ \emph {et~al.}(2017)\citenamefont
  {Van-Damme}, \citenamefont {Schraft}, \citenamefont {Genov}, \citenamefont
  {Sugny}, \citenamefont {Halfmann},\ and\ \citenamefont
  {Gu\'erin}}]{Van-Damme2017}%
  \BibitemOpen
  \bibfield  {author} {\bibinfo {author} {\bibfnamefont {L.}~\bibnamefont
  {Van-Damme}}, \bibinfo {author} {\bibfnamefont {D.}~\bibnamefont {Schraft}},
  \bibinfo {author} {\bibfnamefont {G.~T.}\ \bibnamefont {Genov}}, \bibinfo
  {author} {\bibfnamefont {D.}~\bibnamefont {Sugny}}, \bibinfo {author}
  {\bibfnamefont {T.}~\bibnamefont {Halfmann}},\ and\ \bibinfo {author}
  {\bibfnamefont {S.}~\bibnamefont {Gu\'erin}},\ }\bibfield  {title} {\bibinfo
  {title} {Robust not gate by single-shot-shaped pulses: {D}emonstration of the
  efficiency of the pulses in rephasing atomic coherences},\ }\href
  {https://doi.org/10.1103/PhysRevA.96.022309} {\bibfield  {journal} {\bibinfo
  {journal} {Phys. Rev. A}\ }\textbf {\bibinfo {volume} {96}},\ \bibinfo
  {pages} {022309} (\bibinfo {year} {2017})}\BibitemShut {NoStop}%
\bibitem [{\citenamefont {Levitt}(1986)}]{Levitt1986}%
  \BibitemOpen
  \bibfield  {author} {\bibinfo {author} {\bibfnamefont {M.~H.}\ \bibnamefont
  {Levitt}},\ }\bibfield  {title} {\bibinfo {title} {{Composite pulses}},\
  }\href {https://doi.org/10.1016/0079-6565(86)80005-X} {\bibfield  {journal}
  {\bibinfo  {journal} {Prog. Nucl. Magn. Reson. Spectrosc.}\ }\textbf
  {\bibinfo {volume} {18}},\ \bibinfo {pages} {61} (\bibinfo {year}
  {1986})}\BibitemShut {NoStop}%
\bibitem [{\citenamefont {Torosov}\ \emph {et~al.}(2011)\citenamefont
  {Torosov}, \citenamefont {Gu{\'{e}}rin},\ and\ \citenamefont
  {Vitanov}}]{Torosov2011}%
  \BibitemOpen
  \bibfield  {author} {\bibinfo {author} {\bibfnamefont {B.~T.}\ \bibnamefont
  {Torosov}}, \bibinfo {author} {\bibfnamefont {S.}~\bibnamefont
  {Gu{\'{e}}rin}},\ and\ \bibinfo {author} {\bibfnamefont {N.~V.}\ \bibnamefont
  {Vitanov}},\ }\bibfield  {title} {\bibinfo {title} {{High-fidelity adiabatic
  passage by composite sequences of chirped pulses}},\ }\href
  {https://doi.org/10.1103/PhysRevLett.106.233001} {\bibfield  {journal}
  {\bibinfo  {journal} {Phys. Rev. Lett.}\ }\textbf {\bibinfo {volume} {106}},\
  \bibinfo {pages} {233001} (\bibinfo {year} {2011})}\BibitemShut {NoStop}%
\bibitem [{\citenamefont {Torosov}\ and\ \citenamefont
  {Vitanov}(2011)}]{Torosov2011a}%
  \BibitemOpen
  \bibfield  {author} {\bibinfo {author} {\bibfnamefont {B.~T.}\ \bibnamefont
  {Torosov}}\ and\ \bibinfo {author} {\bibfnamefont {N.~V.}\ \bibnamefont
  {Vitanov}},\ }\bibfield  {title} {\bibinfo {title} {{Smooth composite pulses
  for high-fidelity quantum information processing}},\ }\href
  {https://doi.org/10.1103/PhysRevA.83.053420} {\bibfield  {journal} {\bibinfo
  {journal} {Phys. Rev. A}\ }\textbf {\bibinfo {volume} {83}},\ \bibinfo
  {pages} {053420} (\bibinfo {year} {2011})}\BibitemShut {NoStop}%
\bibitem [{\citenamefont {Genov}\ \emph {et~al.}(2011)\citenamefont {Genov},
  \citenamefont {Torosov},\ and\ \citenamefont {Vitanov}}]{Genov2011}%
  \BibitemOpen
  \bibfield  {author} {\bibinfo {author} {\bibfnamefont {G.~T.}\ \bibnamefont
  {Genov}}, \bibinfo {author} {\bibfnamefont {B.~T.}\ \bibnamefont {Torosov}},\
  and\ \bibinfo {author} {\bibfnamefont {N.~V.}\ \bibnamefont {Vitanov}},\
  }\bibfield  {title} {\bibinfo {title} {{Optimized control of multistate
  quantum systems by composite pulse sequences}},\ }\href
  {https://doi.org/10.1103/PhysRevA.84.063413} {\bibfield  {journal} {\bibinfo
  {journal} {Phys. Rev. A}\ }\textbf {\bibinfo {volume} {84}},\ \bibinfo
  {pages} {063413} (\bibinfo {year} {2011})}\BibitemShut {NoStop}%
\bibitem [{\citenamefont {Genov}\ \emph {et~al.}(2014)\citenamefont {Genov},
  \citenamefont {Schraft}, \citenamefont {Halfmann},\ and\ \citenamefont
  {Vitanov}}]{Genov2014}%
  \BibitemOpen
  \bibfield  {author} {\bibinfo {author} {\bibfnamefont {G.~T.}\ \bibnamefont
  {Genov}}, \bibinfo {author} {\bibfnamefont {D.}~\bibnamefont {Schraft}},
  \bibinfo {author} {\bibfnamefont {T.}~\bibnamefont {Halfmann}},\ and\
  \bibinfo {author} {\bibfnamefont {N.~V.}\ \bibnamefont {Vitanov}},\
  }\bibfield  {title} {\bibinfo {title} {Correction of arbitrary field errors
  in population inversion of quantum systems by universal composite pulses},\
  }\href {https://doi.org/10.1103/PhysRevLett.113.043001} {\bibfield  {journal}
  {\bibinfo  {journal} {Phys. Rev. Lett.}\ }\textbf {\bibinfo {volume} {113}},\
  \bibinfo {pages} {043001} (\bibinfo {year} {2014})}\BibitemShut {NoStop}%
\bibitem [{\citenamefont {Torosov}\ and\ \citenamefont
  {Vitanov}(2013)}]{Torosov2013}%
  \BibitemOpen
  \bibfield  {author} {\bibinfo {author} {\bibfnamefont {B.~T.}\ \bibnamefont
  {Torosov}}\ and\ \bibinfo {author} {\bibfnamefont {N.~V.}\ \bibnamefont
  {Vitanov}},\ }\bibfield  {title} {\bibinfo {title} {Composite stimulated
  {R}aman adiabatic passage},\ }\href
  {https://doi.org/10.1103/PhysRevA.87.043418} {\bibfield  {journal} {\bibinfo
  {journal} {Phys. Rev. A}\ }\textbf {\bibinfo {volume} {87}},\ \bibinfo
  {pages} {043418} (\bibinfo {year} {2013})}\BibitemShut {NoStop}%
\bibitem [{\citenamefont {Bruns}\ \emph {et~al.}(2018)\citenamefont {Bruns},
  \citenamefont {Genov}, \citenamefont {Hain}, \citenamefont {Vitanov},\ and\
  \citenamefont {Halfmann}}]{Bruns2018}%
  \BibitemOpen
  \bibfield  {author} {\bibinfo {author} {\bibfnamefont {A.}~\bibnamefont
  {Bruns}}, \bibinfo {author} {\bibfnamefont {G.~T.}\ \bibnamefont {Genov}},
  \bibinfo {author} {\bibfnamefont {M.}~\bibnamefont {Hain}}, \bibinfo {author}
  {\bibfnamefont {N.~V.}\ \bibnamefont {Vitanov}},\ and\ \bibinfo {author}
  {\bibfnamefont {T.}~\bibnamefont {Halfmann}},\ }\bibfield  {title} {\bibinfo
  {title} {{Experimental demonstration of composite stimulated Raman adiabatic
  passage}},\ }\href {https://doi.org/10.1103/PhysRevA.98.053413} {\bibfield
  {journal} {\bibinfo  {journal} {Phys. Rev. A}\ }\textbf {\bibinfo {volume}
  {98}},\ \bibinfo {pages} {053413} (\bibinfo {year} {2018})}\BibitemShut
  {NoStop}%
\bibitem [{\citenamefont {Demirplak}\ and\ \citenamefont
  {Rice}(2005)}]{Demirplak2005}%
  \BibitemOpen
  \bibfield  {author} {\bibinfo {author} {\bibfnamefont {M.}~\bibnamefont
  {Demirplak}}\ and\ \bibinfo {author} {\bibfnamefont {S.~A.}\ \bibnamefont
  {Rice}},\ }\bibfield  {title} {\bibinfo {title} {{Assisted Adiabatic Passage
  Revisited}},\ }\href {https://doi.org/10.1021/JP040647W} {\bibfield
  {journal} {\bibinfo  {journal} {J. Phys. Chem. B}\ }\textbf {\bibinfo
  {volume} {109}},\ \bibinfo {pages} {6838} (\bibinfo {year}
  {2005})}\BibitemShut {NoStop}%
\bibitem [{\citenamefont {Demirplak}\ and\ \citenamefont
  {Rice}(2008)}]{Demirplak2008}%
  \BibitemOpen
  \bibfield  {author} {\bibinfo {author} {\bibfnamefont {M.}~\bibnamefont
  {Demirplak}}\ and\ \bibinfo {author} {\bibfnamefont {S.~A.}\ \bibnamefont
  {Rice}},\ }\bibfield  {title} {\bibinfo {title} {{On the consistency,
  extremal, and global properties of counterdiabatic fields}},\ }\href
  {https://doi.org/10.1063/1.2992152} {\bibfield  {journal} {\bibinfo
  {journal} {J. Chem. Phys.}\ }\textbf {\bibinfo {volume} {129}},\ \bibinfo
  {pages} {154111} (\bibinfo {year} {2008})}\BibitemShut {NoStop}%
\bibitem [{\citenamefont {Berry}(2009)}]{Berry2009}%
  \BibitemOpen
  \bibfield  {author} {\bibinfo {author} {\bibfnamefont {M.~V.}\ \bibnamefont
  {Berry}},\ }\bibfield  {title} {\bibinfo {title} {{Transitionless quantum
  driving}},\ }\href {https://doi.org/10.1088/1751-8113/42/36/365303}
  {\bibfield  {journal} {\bibinfo  {journal} {J. Phys. A}\ }\textbf {\bibinfo
  {volume} {42}},\ \bibinfo {pages} {365303} (\bibinfo {year}
  {2009})}\BibitemShut {NoStop}%
\bibitem [{\citenamefont {Chen}\ \emph {et~al.}(2010)\citenamefont {Chen},
  \citenamefont {Lizuain}, \citenamefont {Ruschhaupt}, \citenamefont
  {Gu\'ery-Odelin},\ and\ \citenamefont {Muga}}]{Chen2010}%
  \BibitemOpen
  \bibfield  {author} {\bibinfo {author} {\bibfnamefont {X.}~\bibnamefont
  {Chen}}, \bibinfo {author} {\bibfnamefont {I.}~\bibnamefont {Lizuain}},
  \bibinfo {author} {\bibfnamefont {A.}~\bibnamefont {Ruschhaupt}}, \bibinfo
  {author} {\bibfnamefont {D.}~\bibnamefont {Gu\'ery-Odelin}},\ and\ \bibinfo
  {author} {\bibfnamefont {J.~G.}\ \bibnamefont {Muga}},\ }\bibfield  {title}
  {\bibinfo {title} {Shortcut to adiabatic passage in two- and three-level
  atoms},\ }\href {https://doi.org/10.1103/PhysRevLett.105.123003} {\bibfield
  {journal} {\bibinfo  {journal} {Phys. Rev. Lett.}\ }\textbf {\bibinfo
  {volume} {105}},\ \bibinfo {pages} {123003} (\bibinfo {year}
  {2010})}\BibitemShut {NoStop}%
\bibitem [{\citenamefont {Chen}\ \emph {et~al.}(2011)\citenamefont {Chen},
  \citenamefont {Torrontegui},\ and\ \citenamefont {Muga}}]{Chen2011}%
  \BibitemOpen
  \bibfield  {author} {\bibinfo {author} {\bibfnamefont {X.}~\bibnamefont
  {Chen}}, \bibinfo {author} {\bibfnamefont {E.}~\bibnamefont {Torrontegui}},\
  and\ \bibinfo {author} {\bibfnamefont {J.~G.}\ \bibnamefont {Muga}},\
  }\bibfield  {title} {\bibinfo {title} {Lewis\nobreakhyphen {R}iesenfeld
  invariants and transitionless quantum driving},\ }\href
  {https://doi.org/10.1103/PhysRevA.83.062116} {\bibfield  {journal} {\bibinfo
  {journal} {Phys Rev A}\ }\textbf {\bibinfo {volume} {83}},\ \bibinfo {pages}
  {062116} (\bibinfo {year} {2011})}\BibitemShut {NoStop}%
\bibitem [{\citenamefont {Bason}\ \emph {et~al.}(2012)\citenamefont {Bason},
  \citenamefont {Viteau}, \citenamefont {Malossi}, \citenamefont {Huillery},
  \citenamefont {Arimondo}, \citenamefont {Ciampini}, \citenamefont {Fazio},
  \citenamefont {Giovannetti}, \citenamefont {Mannella},\ and\ \citenamefont
  {Morsch}}]{Bason2012}%
  \BibitemOpen
  \bibfield  {author} {\bibinfo {author} {\bibfnamefont {M.~G.}\ \bibnamefont
  {Bason}}, \bibinfo {author} {\bibfnamefont {M.}~\bibnamefont {Viteau}},
  \bibinfo {author} {\bibfnamefont {N.}~\bibnamefont {Malossi}}, \bibinfo
  {author} {\bibfnamefont {P.}~\bibnamefont {Huillery}}, \bibinfo {author}
  {\bibfnamefont {E.}~\bibnamefont {Arimondo}}, \bibinfo {author}
  {\bibfnamefont {D.}~\bibnamefont {Ciampini}}, \bibinfo {author}
  {\bibfnamefont {R.}~\bibnamefont {Fazio}}, \bibinfo {author} {\bibfnamefont
  {V.}~\bibnamefont {Giovannetti}}, \bibinfo {author} {\bibfnamefont
  {R.}~\bibnamefont {Mannella}},\ and\ \bibinfo {author} {\bibfnamefont
  {O.}~\bibnamefont {Morsch}},\ }\bibfield  {title} {\bibinfo {title}
  {{High-fidelity quantum driving}},\ }\href
  {https://doi.org/10.1038/nphys2170} {\bibfield  {journal} {\bibinfo
  {journal} {Nat. Phys.}\ }\textbf {\bibinfo {volume} {8}},\ \bibinfo {pages}
  {147} (\bibinfo {year} {2012})}\BibitemShut {NoStop}%
\bibitem [{\citenamefont {Ruschhaupt}\ \emph {et~al.}(2012)\citenamefont
  {Ruschhaupt}, \citenamefont {Chen}, \citenamefont {Alonso},\ and\
  \citenamefont {Muga}}]{Ruschhaupt2012}%
  \BibitemOpen
  \bibfield  {author} {\bibinfo {author} {\bibfnamefont {A.}~\bibnamefont
  {Ruschhaupt}}, \bibinfo {author} {\bibfnamefont {X.}~\bibnamefont {Chen}},
  \bibinfo {author} {\bibfnamefont {D.}~\bibnamefont {Alonso}},\ and\ \bibinfo
  {author} {\bibfnamefont {J.~G.}\ \bibnamefont {Muga}},\ }\bibfield  {title}
  {\bibinfo {title} {Optimally robust shortcuts to population inversion in
  two-level quantum systems},\ }\href
  {https://doi.org/10.1088/1367-2630/14/9/093040} {\bibfield  {journal}
  {\bibinfo  {journal} {New J. Phys.}\ }\textbf {\bibinfo {volume} {14}},\
  \bibinfo {pages} {093040} (\bibinfo {year} {2012})}\BibitemShut {NoStop}%
\bibitem [{\citenamefont {Baksic}\ \emph {et~al.}(2016)\citenamefont {Baksic},
  \citenamefont {Ribeiro},\ and\ \citenamefont {Clerk}}]{Baksic2016}%
  \BibitemOpen
  \bibfield  {author} {\bibinfo {author} {\bibfnamefont {A.}~\bibnamefont
  {Baksic}}, \bibinfo {author} {\bibfnamefont {H.}~\bibnamefont {Ribeiro}},\
  and\ \bibinfo {author} {\bibfnamefont {A.~A.}\ \bibnamefont {Clerk}},\
  }\bibfield  {title} {\bibinfo {title} {{Speeding up Adiabatic Quantum State
  Transfer by Using Dressed States}},\ }\href
  {https://doi.org/10.1103/PhysRevLett.116.230503} {\bibfield  {journal}
  {\bibinfo  {journal} {Phys. Rev. Lett.}\ }\textbf {\bibinfo {volume} {116}},\
  \bibinfo {pages} {230503} (\bibinfo {year} {2016})}\BibitemShut {NoStop}%
\bibitem [{\citenamefont {Li}\ and\ \citenamefont {Chen}(2016)}]{Li2016}%
  \BibitemOpen
  \bibfield  {author} {\bibinfo {author} {\bibfnamefont {Y.~C.}\ \bibnamefont
  {Li}}\ and\ \bibinfo {author} {\bibfnamefont {X.}~\bibnamefont {Chen}},\
  }\bibfield  {title} {\bibinfo {title} {{Shortcut to adiabatic population
  transfer in quantum three-level systems: Effective two-level problems and
  feasible counterdiabatic driving}},\ }\href
  {https://doi.org/10.1103/PhysRevA.94.063411} {\bibfield  {journal} {\bibinfo
  {journal} {Phys. Rev. A}\ }\textbf {\bibinfo {volume} {94}},\ \bibinfo
  {pages} {063411} (\bibinfo {year} {2016})}\BibitemShut {NoStop}%
\bibitem [{\citenamefont {Ban}\ \emph {et~al.}(2018)\citenamefont {Ban},
  \citenamefont {Jiang}, \citenamefont {Li}, \citenamefont {Wang},\ and\
  \citenamefont {Chen}}]{Ban2018}%
  \BibitemOpen
  \bibfield  {author} {\bibinfo {author} {\bibfnamefont {Y.}~\bibnamefont
  {Ban}}, \bibinfo {author} {\bibfnamefont {L.~X.}\ \bibnamefont {Jiang}},
  \bibinfo {author} {\bibfnamefont {Y.~C.}\ \bibnamefont {Li}}, \bibinfo
  {author} {\bibfnamefont {L.~J.}\ \bibnamefont {Wang}},\ and\ \bibinfo
  {author} {\bibfnamefont {X.}~\bibnamefont {Chen}},\ }\bibfield  {title}
  {\bibinfo {title} {{Fast creation and transfer of coherence in triple quantum
  dots by using shortcuts to adiabaticity}},\ }\href
  {https://doi.org/10.1364/OE.26.031137} {\bibfield  {journal} {\bibinfo
  {journal} {Opt. Express}\ }\textbf {\bibinfo {volume} {26}},\ \bibinfo
  {pages} {31137} (\bibinfo {year} {2018})}\BibitemShut {NoStop}%
\bibitem [{\citenamefont {Lewis}\ and\ \citenamefont
  {Riesenfeld}(1969)}]{Lewis1969}%
  \BibitemOpen
  \bibfield  {author} {\bibinfo {author} {\bibfnamefont {H.~R.}\ \bibnamefont
  {Lewis}}\ and\ \bibinfo {author} {\bibfnamefont {W.~B.}\ \bibnamefont
  {Riesenfeld}},\ }\bibfield  {title} {\bibinfo {title} {{An Exact Quantum
  Theory of the Time-Dependent Harmonic Oscillator and of a Charged Particle in
  a Time-Dependent Electromagnetic Field}},\ }\href
  {https://doi.org/10.1063/1.1664991} {\bibfield  {journal} {\bibinfo
  {journal} {J. Math. Phys.}\ }\textbf {\bibinfo {volume} {10}},\ \bibinfo
  {pages} {1458} (\bibinfo {year} {1969})}\BibitemShut {NoStop}%
\bibitem [{\citenamefont {Lai}\ \emph {et~al.}(1996{\natexlab{a}})\citenamefont
  {Lai}, \citenamefont {Liang}, \citenamefont {M{\"{u}}ller-Kirsten},\ and\
  \citenamefont {Zhou}}]{Lai1996}%
  \BibitemOpen
  \bibfield  {author} {\bibinfo {author} {\bibfnamefont {Y.~Z.}\ \bibnamefont
  {Lai}}, \bibinfo {author} {\bibfnamefont {J.~Q.}\ \bibnamefont {Liang}},
  \bibinfo {author} {\bibfnamefont {H.~J.~W.}\ \bibnamefont
  {M{\"{u}}ller-Kirsten}},\ and\ \bibinfo {author} {\bibfnamefont {J.~G.}\
  \bibnamefont {Zhou}},\ }\bibfield  {title} {\bibinfo {title} {{Time evolution
  of quantum systems with time-dependent Hamiltonian and the invariant
  Hermitian operator}},\ }\href {https://doi.org/10.1088/0305-4470/29/8/024}
  {\bibfield  {journal} {\bibinfo  {journal} {J. Phys. A}\ }\textbf {\bibinfo
  {volume} {29}},\ \bibinfo {pages} {1773} (\bibinfo {year}
  {1996}{\natexlab{a}})}\BibitemShut {NoStop}%
\bibitem [{\citenamefont {Lai}\ \emph {et~al.}(1996{\natexlab{b}})\citenamefont
  {Lai}, \citenamefont {Liang}, \citenamefont {M{\"{u}}ller-Kirsten},\ and\
  \citenamefont {Zhou}}]{Lai1996a}%
  \BibitemOpen
  \bibfield  {author} {\bibinfo {author} {\bibfnamefont {Y.~Z.}\ \bibnamefont
  {Lai}}, \bibinfo {author} {\bibfnamefont {J.~Q.}\ \bibnamefont {Liang}},
  \bibinfo {author} {\bibfnamefont {H.~J.~W.}\ \bibnamefont
  {M{\"{u}}ller-Kirsten}},\ and\ \bibinfo {author} {\bibfnamefont {J.~G.}\
  \bibnamefont {Zhou}},\ }\bibfield  {title} {\bibinfo {title} {Time-dependent
  quantum systems and the invariant {H}ermitian operator},\ }\href
  {https://doi.org/10.1103/PhysRevA.53.3691} {\bibfield  {journal} {\bibinfo
  {journal} {Phys. Rev. A}\ }\textbf {\bibinfo {volume} {53}},\ \bibinfo
  {pages} {3691} (\bibinfo {year} {1996}{\natexlab{b}})}\BibitemShut {NoStop}%
\bibitem [{\citenamefont {Gell-Mann}(1962)}]{GellMann1962}%
  \BibitemOpen
  \bibfield  {author} {\bibinfo {author} {\bibfnamefont {M.}~\bibnamefont
  {Gell-Mann}},\ }\bibfield  {title} {\bibinfo {title} {{Symmetries of Baryons
  and Mesons}},\ }\href {https://doi.org/10.1103/PhysRev.125.1067} {\bibfield
  {journal} {\bibinfo  {journal} {Phys. Rev.}\ }\textbf {\bibinfo {volume}
  {125}},\ \bibinfo {pages} {1067} (\bibinfo {year} {1962})}\BibitemShut
  {NoStop}%
\bibitem [{\citenamefont {Carroll}\ and\ \citenamefont
  {Hioe}(1988)}]{Carroll1988}%
  \BibitemOpen
  \bibfield  {author} {\bibinfo {author} {\bibfnamefont {C.~E.}\ \bibnamefont
  {Carroll}}\ and\ \bibinfo {author} {\bibfnamefont {F.~T.}\ \bibnamefont
  {Hioe}},\ }\bibfield  {title} {\bibinfo {title} {{Three-state systems driven
  by resonant optical pulses of different shapes}},\ }\href
  {https://doi.org/10.1364/JOSAB.5.001335} {\bibfield  {journal} {\bibinfo
  {journal} {J. Opt. Soc. Am. B}\ }\textbf {\bibinfo {volume} {5}},\ \bibinfo
  {pages} {1335} (\bibinfo {year} {1988})}\BibitemShut {NoStop}%
\bibitem [{\citenamefont {Bergmann}\ \emph {et~al.}(1998)\citenamefont
  {Bergmann}, \citenamefont {Theuer},\ and\ \citenamefont
  {Shore}}]{Bergmann1998}%
  \BibitemOpen
  \bibfield  {author} {\bibinfo {author} {\bibfnamefont {K.}~\bibnamefont
  {Bergmann}}, \bibinfo {author} {\bibfnamefont {H.}~\bibnamefont {Theuer}},\
  and\ \bibinfo {author} {\bibfnamefont {B.~W.}\ \bibnamefont {Shore}},\
  }\bibfield  {title} {\bibinfo {title} {Coherent population transfer among
  quantum states of atoms and molecules},\ }\href
  {https://doi.org/10.1103/RevModPhys.70.1003} {\bibfield  {journal} {\bibinfo
  {journal} {Rev. Mod. Phys.}\ }\textbf {\bibinfo {volume} {70}},\ \bibinfo
  {pages} {1003} (\bibinfo {year} {1998})}\BibitemShut {NoStop}%
\bibitem [{\citenamefont {Rousseaux}\ \emph {et~al.}(2013)\citenamefont
  {Rousseaux}, \citenamefont {Gu\'erin},\ and\ \citenamefont
  {Vitanov}}]{Rousseaux2013}%
  \BibitemOpen
  \bibfield  {author} {\bibinfo {author} {\bibfnamefont {B.}~\bibnamefont
  {Rousseaux}}, \bibinfo {author} {\bibfnamefont {S.}~\bibnamefont
  {Gu\'erin}},\ and\ \bibinfo {author} {\bibfnamefont {N.~V.}\ \bibnamefont
  {Vitanov}},\ }\bibfield  {title} {\bibinfo {title} {Arbitrary qudit gates by
  adiabatic passage},\ }\href {https://doi.org/10.1103/PhysRevA.87.032328}
  {\bibfield  {journal} {\bibinfo  {journal} {Phys. Rev. A}\ }\textbf {\bibinfo
  {volume} {87}},\ \bibinfo {pages} {032328} (\bibinfo {year}
  {2013})}\BibitemShut {NoStop}%
\bibitem [{\citenamefont {Hioe}(1985)}]{Hioe1985}%
  \BibitemOpen
  \bibfield  {author} {\bibinfo {author} {\bibfnamefont {F.~T.}\ \bibnamefont
  {Hioe}},\ }\bibfield  {title} {\bibinfo {title} {{Gell-Mann dynamic symmetry
  for N-level quantum systems}},\ }\href
  {https://doi.org/10.1103/PhysRevA.32.2824} {\bibfield  {journal} {\bibinfo
  {journal} {Phys. Rev. A}\ }\textbf {\bibinfo {volume} {32}},\ \bibinfo
  {pages} {2824} (\bibinfo {year} {1985})}\BibitemShut {NoStop}%
\bibitem [{\citenamefont {Hioe}(1987)}]{Hioe1987}%
  \BibitemOpen
  \bibfield  {author} {\bibinfo {author} {\bibfnamefont {F.~T.}\ \bibnamefont
  {Hioe}},\ }\bibfield  {title} {\bibinfo {title} {{N-level quantum systems
  with SU(2) dynamic symmetry}},\ }\href
  {https://doi.org/10.1364/JOSAB.4.001327} {\bibfield  {journal} {\bibinfo
  {journal} {J.~Opt.~Soc.~Am.~B}\ }\textbf {\bibinfo {volume} {4}},\ \bibinfo
  {pages} {1327} (\bibinfo {year} {1987})}\BibitemShut {NoStop}%
\bibitem [{\citenamefont {Hioe}(1988)}]{Hioe1988}%
  \BibitemOpen
  \bibfield  {author} {\bibinfo {author} {\bibfnamefont {F.~T.}\ \bibnamefont
  {Hioe}},\ }\bibfield  {title} {\bibinfo {title} {{N-level quantum systems
  with Gell-Mann dynamic symmetry}},\ }\href
  {https://doi.org/10.1364/JOSAB.5.000859} {\bibfield  {journal} {\bibinfo
  {journal} {J. Opt. Soc. Am. B}\ }\textbf {\bibinfo {volume} {5}},\ \bibinfo
  {pages} {859} (\bibinfo {year} {1988})}\BibitemShut {NoStop}%
\bibitem [{\citenamefont {G\"ung\"ord\"u}\ \emph {et~al.}(2012)\citenamefont
  {G\"ung\"ord\"u}, \citenamefont {Wan}, \citenamefont {Fasihi},\ and\
  \citenamefont {Nakahara}}]{Guengoerdue2012}%
  \BibitemOpen
  \bibfield  {author} {\bibinfo {author} {\bibfnamefont {U.}~\bibnamefont
  {G\"ung\"ord\"u}}, \bibinfo {author} {\bibfnamefont {Y.}~\bibnamefont {Wan}},
  \bibinfo {author} {\bibfnamefont {M.~A.}\ \bibnamefont {Fasihi}},\ and\
  \bibinfo {author} {\bibfnamefont {M.}~\bibnamefont {Nakahara}},\ }\bibfield
  {title} {\bibinfo {title} {Dynamical invariants for quantum control of
  four-level systems},\ }\href {https://doi.org/10.1103/PhysRevA.86.062312}
  {\bibfield  {journal} {\bibinfo  {journal} {Phys. Rev. A}\ }\textbf {\bibinfo
  {volume} {86}},\ \bibinfo {pages} {062312} (\bibinfo {year}
  {2012})}\BibitemShut {NoStop}%
\bibitem [{\citenamefont {Shore}(2014)}]{Shore2013}%
  \BibitemOpen
  \bibfield  {author} {\bibinfo {author} {\bibfnamefont {B.~W.}\ \bibnamefont
  {Shore}},\ }\bibfield  {title} {\bibinfo {title} {Two-state behavior in
  {N}-state quantum systems: The {M}orris-{S}hore transformation reviewed},\
  }\href {https://doi.org/10.1080/09500340.2013.837205} {\bibfield  {journal}
  {\bibinfo  {journal} {J. Mod. Opt}\ }\textbf {\bibinfo {volume} {61}},\
  \bibinfo {pages} {787} (\bibinfo {year} {2014})}\BibitemShut {NoStop}%
\end{thebibliography}
%

\end{document}